\documentclass[twocolumn,showpacs,preprintnumbers,amsmath,amssymb,floatfix]{revtex4}

\usepackage{graphicx}

\def\dirfig{}

\begin{document}


\title{Hydrogen Release from Sodium Alanate\\
  Observed by Time-resolved Neutron Backscattering}

\author{Aline L\'eon}
 \affiliation{Karlsruhe Institute of Technology,
Institute of Nanotechnology,
Hermann-von-Helmholtz-Platz 1,
76344 Eggenstein-Leopoldshafen, Germany}
 \email{aline.leon@kit.edu}
\author{Joachim Wuttke}
 \affiliation{J\"ulich Centre for Neutron Science JCNS,
Forschungszentrum J\"ulich GmbH,
Outstation at FRM II,
Lichtenbergstr.~1, 85747 Garching, Germany}
 \email{j.wuttke@fz-juelich.de}

\date{\today}

\begin{abstract}
Innermolecular motion in Na$_3$AlH$_6$ gives rise to
a Lorentzian spectrum
with a wavenumber-independent width of about 1~$\mu$eV at 180$^\circ$C,
which is probably due to rotation of AlH$_6$ tetrahedra.
There is no such quasielastic line in NaAlH$_4$ or NaH. 
Based on this finding,
time-resolved measurements on the neutron backscattering spectrometer SPHERES
were used to monitor the decomposition kinetics
of sodium alanate, NaAlH$_4$ $\to$ Na$_3$AlH$_6$ $\to$ NaH.
Both reaction steps were found to be accelerated by autocatalysis,
most likely at the surfaces of Na$_3$AlH$_6$ and NaH crystallites.
\end{abstract}

\pacs{88.30.R-,82.30.Lp,63.50.-x,61.05.F-}


\maketitle

\section{Introduction}

Sodium alanate NaAlH$_4$ is
widely studied as a model system for hydrogen storage
\cite{ScBF04,OrNE07,EbFS09}.
While hydrogen exchange in pure bulk alanate is not reversible,
 hydrogen desorption and absorption kinetics
has been improved decisively by doping \cite{BoSc97,Ant03}
and nanostructuring \cite{ZaZa99,HuBo99,KaWC07}.
Further improvement of the hydriding/dehydriding kinetics
requires a detailed understanding of the reaction mechanism.

In this work,
we introduce time-resolved neutron backscattering as a new tool
to monitor the solid-state reactions that are responsible for
hydrogen release.
Measurements are carried out on pure NaAlH$_4$
and its decomposition products Na$_3$AlH$_6$ and NaH;
one explorative experiment has been undertaken on material doped with TiCl$_3$,
cycled under hydrogen, and quenched in the hydrogenated state.

\subsection{The Hydriding/Dehydriding Reactions}\label{SIpure}

Hydrogen storage in sodium alanate
is based on the two reactions
\begin{eqnarray}
    \mbox{NaAlH$_4$} &\leftrightarrow& \label{Estep1} \textstyle
 \frac{1}{3} \mbox{Na$_3$AlH$_6$} + \frac{2}{3} \mbox{Al} + \mbox{H$_2$},\\[2ex]
       \mbox{Na$_3$AlH$_6$} &\leftrightarrow& \label{Estep2} \textstyle
    3 \mbox{NaH} + \mbox{Al} + \frac{3}{2} \mbox{H$_2$}.
\end{eqnarray}
A third reaction, the decomposition of NaH, has no practical importance,
because it only occurs above 425$^\circ$C \cite{BoBM00}.
Therefore,
a theoretically reversible capacity of 5.6 wt.\% hydrogen is available. 
The equilibrium temperatures of (\ref{Estep1}) and~(\ref{Estep2})
depend only little on doping, 
but much on pressure \cite{LuGr04}.
At 1~bar, they are about 30$^\circ$C and 110$^\circ$C, respectively,
with uncertainties of several degrees \cite[Fig.~2]{LeJS06}.
A thermodynamic argument suggests that at
particle sizes below 50~nm,
NaAlH$_4$ may decompose in a single step into  NaH, Al, and H$_2$ \cite{MuCe10}.
On the other hand,
time-resolved neutron diffraction during
hydrogen release from NaAlH$_4$ with a particle size of 110~nm
clearly shows a transitory state rich in Na$_3$AlH$_6$ \cite{SiEH07}.

In undoped bulk NaAlH$_4$, reaction rates for hydrogen release are very low,
unless the temperature approaches the melting point 
for which values between 175$^\circ$C and 190$^\circ$C are
reported \cite{ClBC80,JeGr01,KiFi04,SiEH07}.
In pure ball-milled NaAlH$_4$ at 170$^\circ$,
it takes at least 8~h to complete the desorption step~(\ref{Estep1}).
When doped with Ti clusters, for example, this step takes only 300~s
at 150$^\circ$C \cite{KiFi04}.

From studies on doped material,
it can be concluded that the reactions (\ref{Estep1}),~(\ref{Estep2})
take place at the boundary between Al particles
and NaH, Na$_3$AlH$_6$, or NaAlH$_4$ phases \cite{FeKG04}
and that the rates of these reactions are limited
by a nucleation/growth process \cite{KiSS03}
or/and
by the transport of heavy, Al- or/and Na-based species \cite{ScBF04,FiCK05}.
This is supported by H/D exchange experiments
showing that the rate-limiting step of the
hydriding/dehydriding process is neither H$_2$ dissociation
nor H diffusion \cite{BeSF06,LoFi07}.

Quite different mechanisms have been proposed 
to explain the effect of titanium doping:
(i) Ti is a surface catalyst \cite{ChMu05,BeSF06,Veg06,DaMa10};
(ii) Ti creates Na$^+$ vacancies in the bulk structure,
thereby facilitating hydrogen diffusion \cite{SuKT02,MoLA05,SiEH07}
or/and the transport of heavier species \cite{FiCK05,SiEH07};
(iii) Ti weakens the Al-H bond, promoting the removal of H$_2$
\cite{BaWu06};
(iv) TiCl$_3$ is a grain refiner preventing growth of Al and NaH particles
\cite{SiEH07};
(v) Ti is an initiator of Al nucleation \cite{DaMa10}.

\subsection{Quasielastic Neutron Scattering Studies}

We are aware of two previous quasielastic neutron scattering measurements.
One measurement using a triple-axis spectrometer
on NaAlH$_4$ at 150$^\circ$C indicates
that a small fraction ($<10\%$) of H atoms
participates in a fast localized process 
(almost $q$ independent Lorentzian halfwidth $\Gamma\simeq70$~$\mu$eV,
corresponding to a characteristic time 
of $\tau=\hbar/\Gamma\simeq10$~ps) \cite{AnHV05}.
Doped and undoped NaAlH$_4$ and Na$_3$AlH$_6$ 
were investigated using the 
J\"ulich backscattering spectrometer BSS \cite{ShVJ07,VoSJ07}.
In doped NaAlH$_4$ at 117$^\circ$C,
about 0.5\% of the hydrogen were found to be mobile,
too little
to make reliable statements about its motion \cite{ShVJ07}.
In doped Na$_3$AlH$_6$ at 77$^\circ$C
and in undoped NaAlH$_4$ at 117$^\circ$C,
some quasielastic broadening was observed
although the Lorentzian width $\Gamma$ was considerably smaller than
the resolution fwhm of 0.8~$\mu$eV \cite{VoSJ07}.
As no spectral fits were shown,
it is not possible to assess the conclusion of broadening
being due to jump diffusion.

In the meantime, 
Forschungszentrum J\"ulich has replaced the BSS
by the new backscattering spectrometer SPHERES \cite{spheres}.
Compared to BSS, SPHERES has much higher count rates,
better resolution (0.65~$\mu$eV), a wider dynamic range,
and a better signal-to-noise ratio.
It is now possible to measure a meaningful spectrum
within a fraction of an hour.
This enables us to observe microscopic dynamics
during hydrogen release in real time.

In this work, we will use neutron backscattering
to monitor the reactions (\ref{Estep1}) and (\ref{Estep2}).
Using the quasielastic amplitude as a proxy
for the amount of Na$_3$AlH$_6$ in the sample,
we will show how this compound builds up in the first reaction step
and disappears in the slower second step.
Our results indicate that both reaction steps,
after starting very slowly,
are accelerated significantly when autocatalysis on 
the surface of Na$_3$AlH$_6$ and NaH crystallites sets in.
Preliminary data for TiCl$_3$-doped NaAlH$_4$ suggest
that the dopant acts like a nucleation center,
accelerating the initial, non-autocatalytic phases of both reactions.

\section{Sample Preparation}

In this paper, we report on four sample materials:
NaH, Na$_3$AlH$_6$, NaAlH$_4$, and NaAlH$_4$(a8a).
All samples were prepared in an argon-filled glove box 
 equipped with a recirculation system to keep the water
 and oxygen concentrations below 1 ppm.
The raw chemicals NaH (95\%, Sigma Aldrich),
 NaAlH$_4$ (96\%, Albemarle), and TiCl$_3$ (99.999\%, Sigma Aldrich)
 were used as received.
Mechanical milling was carried out in a Fritsch P6 planetary
 mixer/mill at a rotation speed of 600 rpm
using a silicon nitride vial and balls
with a ball to powder weight ratio of about 20:1. The vial
was filled and sealed in the glove box under argon atmosphere.

The samples were prepared as follows:

NaH: Powder as received.

Na$_3$AlH$_6$: Obtained by mechanical alloying of NaH
and  NaAlH$_4$ at a molar ratio of 2 to 1. 
The powder was milled for 20 h in an
argon atmosphere as described in Refs.~\cite{ZaZa99,HuBo99}.
The product was analyzed by X-ray diffraction.

Pure NaAlH$_4$: Powder as received, ball-milled for 30 minutes.

NaAlH$_4$ (a8a): Doped with TiCl$_3$,
cycled eight times under hydrogen, and quenched in
the hydrogenated state.
This sample was obtained by ball milling
 2~g of NaAlH$_4$ and 285~mg of TiCl$_3$,
resulting in 5~mol.\% Ti doping on the basis of TiCl$_3$.
In order to avoid any increase in the temperature,
the 3 hours of milling were divided
into 30 minutes of milling and 10 minutes pause with 5 repetitions.
Directly after milling, the sample was first dehydrogenated,
then cycled eight times under hydrogen with hydrogen absorption
(100 $^\circ$C, 100~bar) and desorption (150 $^\circ$C, 0.3~bar) conditions,
and finally quenched in the hydrogenated state.
Absorption and desorption of hydrogen
were carried out in a carefully calibrated modified Sieverts apparatus.
A more detailed description of the apparatus,
the reactor,
and the absorption/desorption procedure can be found elsewhere
\cite{FiFK03,KiFi04}.

Unfortunately,
lack of beamtime prevented us from measuring 
another doped sample, freshly prepared without hydrogen cycling.

The samples were filled into flat, top-loading $30\times40\times0.5$~mm$^3$
 Al cells and sealed with Al wire in the glove box.
In the course of this study,
we were concerned about the sample material,
 while releasing hydrogen,
 possibly reacting with the cell.
Therefore, we performed an additional measurement
with NaAlH$_4$ powder filled into a pocket made of Ag foil
 to isolate the sample from the Al walls.
Direct comparison of the neutron spectra revealed no difference.

\section{Measurements and Data Analysis}

Since the samples are rich in hydrogen,
the neutron scattering cross section is dominated by
incoherent scattering by hydrogen.

Measurements were performed using the
neutron backscattering spectrometer SPHERES
 of the J\"ulich Centre for Neutron Science (JCNS)
 at the neutron source FRM~II
 (Forschungs-Neutronenquelle Heinz Maier-Leibnitz) in 
Garching, Germany \cite{spheres}.
To obtain reasonable statistics within short time slices
 we did not use the instrument's full energy range of $\pm31$~$\mu$eV;
instead, the Doppler velocity amplitude was set to 1.3~m/s,
giving access to a window of $\pm8.6$~$\mu$eV.

The sample was mounted in a cryo\-fur\-nace
 and oriented at 135$^\circ$.
In this standard geometry,
self-shielding within the cell limits the range of useable 
 scattering angles to $2\theta\lesssim125^\circ$.
This results in nine large-angle detectors with
 wavenumbers $q=0.6\ldots1.8$~\AA$^{-1}$.
The small-angle detectors
 at $q=0.25\ldots0.46$~\AA$^{-1}$ are not in exact backscattering
 and therefore have comparatively bad resolution.

The raw data reduction was carried out with SLAW \cite{sla1+},
 and further analysis was performed with FRIDA \cite{fri1+}.
Spectra were normalized to the integral intensity at about room temperature.
The room temperature measurements also served as resolution function.

Deviating from common practice,
we did \textit{not} subtract an empty-cell measurement.
Instead, cell scattering was taken into account in the fitting procedure.
In the energy range of SPHERES,
scattering from Al just consists in a weak elastic line.
Besides, there is a flat background
due to various imperfections of the instrument.
This background was determined
 from the baseline of the resolution spectrum,
and included as a fixed component in all fit models.
In fitting,
squared deviations were weighed with the reciprocal standard deviation,
and theoretical expressions are convoluted with the measured resolution
function (minus the aforementioned baseline).

The data were analyzed in an heuristic and iterative way,
trying different procedures and testing different models
before a consistent overall picture emerged.
In the following sections,
 data analysis will be decoupled from interpretation.
In Sect.~\ref{Sspec}, spectral line shapes are analyzed,
and simple fit functions are introduced.
In Sect.~\ref{Skin}, fits are used to
extract information from time-resolved measurements.
Finally, in Sect.~\ref{Sphys},
the observations are interpreted physically.

\section{Spectra}\label{Sspec}

The simplest fit function that works for the entire data set
is
\begin{equation}\label{ESqw}
  S(q,\omega)
  = b + a_\delta \delta(\omega) + \sum_i a_i(q) {\cal L}(\omega,\Gamma_i).
\end{equation}
It consists of a flat background,
an elastic $\delta$ line,
and zero, one, or two Lorentzians
\begin{equation}\label{ECL}
  {\cal L}(\omega;\Gamma) = \frac{1}{\pi}\frac{\Gamma}{\omega^2+\Gamma^2}.
\end{equation}
The flat background is kept constant
at the value determined at room temperature,
except when the overall scattering decreases with time
because of hydrogen desorption.

Unconstrained fits indicate that the Lorentzian linewidths 
do not vary systematically with~$q$, whereas the amplitudes do.
The amplitudes of the Lorentzians increase with $q$ 
before they reach a plateau at about 1~\AA$^{-1}$.
Therefore, in our final analysis of spectral lineshapes
and time series, we average $S(q,\omega)$ over
the $q$ range 0.9\ldots1.8~\AA$^{-1}$
to obtain a strong quasielastic signal with good statistics.

\subsection{\boldmath N\lowercase{a}H}

\begin{figure}[thb]
 \centerline{\includegraphics*[width=.7\columnwidth]{\dirfig 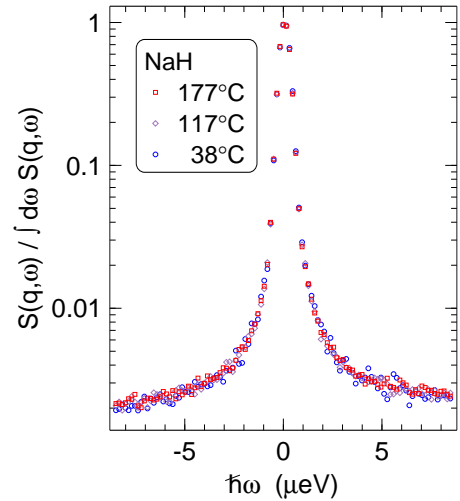}}
 \caption{Neutron scattering spectra of NaH
measured using the backscattering spectrometer SPHERES.
Here and in the following figures,
spectral data are averaged over the $q$~range
$0.9\ldots1.8$~\AA$^{-1}$.
Only in this figure, spectra are normalized to unit area to account for
the temperature dependence of elastic scattering.
There is no temperature-dependent quasielastic scattering;
the spectra represent just the instrumental resolution function.}
\label{Fnah}
\end{figure}

Spectra of the final reaction product NaH
were measured at 38, 117, and 177$^\circ$C.
In Fig.~\ref{Fnah}, these three spectra have been rescaled to account for
the temperature dependence of the elastic scattering
(the Debye-Waller factor, or more precisely the Lamb-M\"ossbauer factor,
since we are talking about self correlations measured by incoherent scattering).
As a result, the spectra coincide perfectly,
demonstrating the absence of quasielastic broadening.

\subsection{\boldmath N\lowercase{a}$_3$A\lowercase{l}H$_6$}

The richest set of spectra was obtained
for the intermediate reaction product Na$_3$AlH$_6$.
Spectra of a freshly prepared sample were measured up to 177$^\circ$C,
where decomposition set in.
Kinetic aspects will be discussed below (Sect.~\ref{Skin}).

\begin{figure}[thb]
 \centerline{\includegraphics*[width=.7\columnwidth]{\dirfig 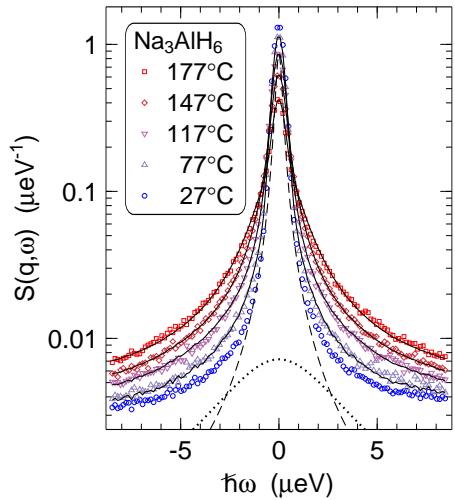}}
 \caption{Neutron scattering spectra of Na$_3$AlH$_6$.
 The 27$^\circ$C data represent the resolution function.
 The other data are fitted according to Eq.~(\ref{ESqw})
 with a delta line and two Lorentzians
  (here and in the following figures, all spectral fits are understood to
  be convoluted with the measured resolution).
 The dashed (dotted) line shows the Lorentzian 1 (2) for 117$^\circ$C
  separately and in unconvoluted form.}
 \label{Fna2-sqw}
\end{figure}

Two Lorentzians are needed to describe the quasielastic scattering.
To reduce the number of free parameters
and to avoid unwanted degeneracies 
(discussed recently in another SPHERES data analysis \cite{DoBG10}),
amplitudes are fixed at temperature-independent values
$a_1=0.5$ and $a_2=0.07$.
Excellent fits are obtained, as shown in Fig.~\ref{Fna2-sqw}.

\begin{figure}[thb]
 \centerline{\includegraphics*[width=.75\columnwidth]{\dirfig 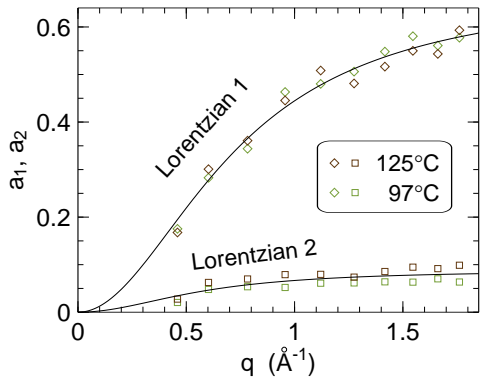}}
 \caption{Wavenumber dependence of the amplitudes $a_{1,2}$ 
   of the two Lorentzian components found 
when fitting the Na$_3$AlH$_6$ spectra with Eq.~(\ref{ESqw}).
 Solid lines: Fits with Eq.~(\ref{Eq2sat}).}
 \label{Fna2-q2a}
\end{figure}

To investigate the $q$ dependence of the amplitudes $a_{1,2}(q)$,
we impose the linewidths $\Gamma_{1,2}(T)$ as 
determined from the fits to the $q$ averaged spectra.
Results are shown in Fig.~\ref{Fna2-q2a}.
As anticipated,
the $a_{1,2}(q)$ are temperature-independent within experimental accuracy.
Their $q$ dependence is fitted reasonably well
by the simple expression
\begin{equation}\label{Eq2sat}
    a_{1,2}(q) = a^\infty_{1,2} \frac{q^2}{q^2+\kappa^2}
\end{equation}
that interpolates between a $q^2$ dependence for $q\ll\kappa$ and
a constant asymptote for $q\gg\kappa$.

\begin{figure}[thb]
 \centerline{\includegraphics*[width=.75\columnwidth]{\dirfig 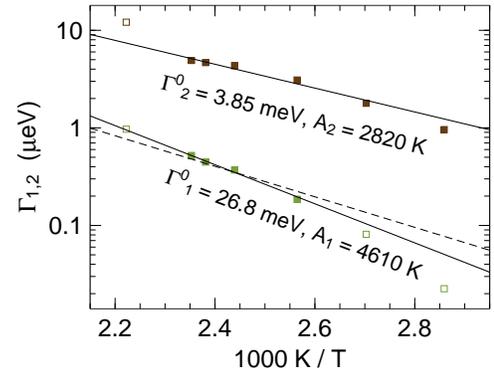}}
 \caption{Linewidths $\Gamma_{1,2}(T)$ of the two Lorentzians used to fit
 the spectra of Na$_3$AlH$_6$.
 Only data shown as full symbols have been used to determine
 the Arrhenius laws shown as straight lines.
 The dashed line is obtained
 when proton NMR data \cite{VeBE09} are taken into account
 (see Fig.~\ref{Fna2+nmr}; ${\Gamma^0_1}'=2.3$~meV, ${A_1}'=3600$~K).}
 \label{Fna2-arr}
\end{figure}

Fig.~\ref{Fna2-arr} shows that
the temperature dependence of the linewidths $\Gamma_{1,2}(T)$
is compatible with an Arrhenius law $\Gamma_i=\Gamma^0_i\exp(-A_i/T)$.
The activation energies are $A_1=38$~kJ/mol and $A_2=23$~kJ/mol;
the prefactors $\Gamma^0_i$ are indicated in the figure.
In estimating these parameters,
some data points were excluded:
linewidths below 0.1~$\mu$eV were dropped, because they are too far
 below the instrumental resolution;
 data at 177$^\circ$C had to be excluded,
 because the scattering signal was not stationary, so that the
 imposition of a fixed Lorentzian amplitude would not have been justified.

\subsection{\boldmath N\lowercase{a}A\lowercase{l}H$_4$}

\begin{figure}[thb]
 \centerline{\includegraphics*[width=.7\columnwidth]{\dirfig 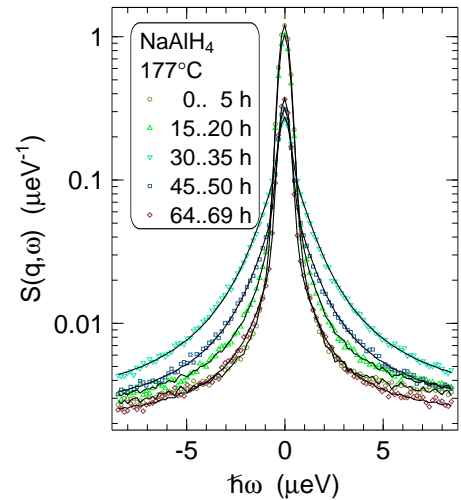}}
 \caption{Neutron scattering spectra of NaAlH$_4$ 
 for various time intervals, after heating to 177$^\circ$C.}
 \label{Fna4-sqw}
\end{figure}

Time-resolved measurements were performed with
four fresh NaAlH$_4$ samples.
After measuring the resolution at room temperature,
the sample was quickly heated to 170, 180, or 185$^\circ$C,
and kept there at constant temperature for 2 days or more.
Initially, no quasielastic scattering is observed.
It takes several hours before a quasielastic signal appears
above the wings of the resolution function.
The signal continues to grow, until a maximum is reached
about 30~h after the start of the measurement.
Then, the quasielastic intensity decreases 
and about 65~h after the start of the measurement
we are left with purely elastic scattering,
which however is only half as strong as in the beginning.

To prepare for the detailed investigation of this kinetics in Sect.~\ref{Skin},
we need to characterise the quasielastic lineshapes.
To do so, the data are averaged over time slots of 5~h
(Fig.~\ref{Fna4-sqw}).
The minimal fit function that describes the entire data set
consists of a flat background, a $\delta$ line, and one Lorentzian.
In contrast to the above analysis of Na$_3$AlH$_6$,
the background must not be kept constant,
because the total scattering intensity decreases during
the experiment.

The Lorentzian linewidth at 180$^\circ$C and for large~$q$ 
is about 1~$\mu$eV.
This is perfectly compatible with the Arrhenius law number~1
of Fig.~\ref{Fna2-arr}.
The slightly smaller and larger linewidths at 170 and 185$^\circ$C
are also compatible with this Arrhenius law.
We therefore attribute the quasielastic scattering observed
during the decomposition of NaAlH$_4$ 
entirely to the Lorentzian~1 that is dominant in Na$_3$AlH$_6$.
In contrast, in the entire NaAlH$_4$ time series we find
 no trace of Lorentzian~2.

\subsection{\boldmath
    N\lowercase{a}A\lowercase{l}H$_4$ (a8a)}

\begin{figure}[thb]
 \centerline{\includegraphics*[width=.7\columnwidth]{\dirfig 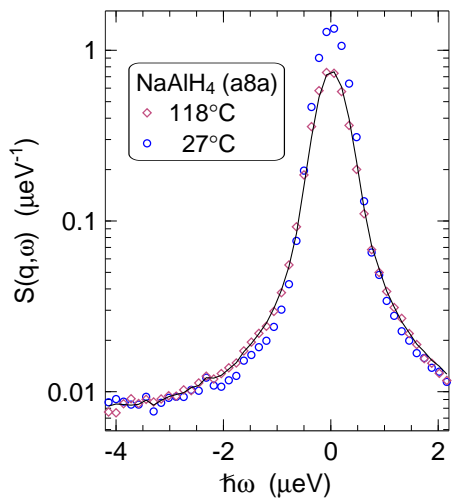}}
 \caption{Neutron scattering spectra of the TiCl$_3$-doped
NaAlH$_4$ (a8a) sample.
 A reduced $\omega$ scale was chosen to emphasize the
 weak quasielastic scattering.}
 \label{Fti4-sqw}
\end{figure}

Due to a lack of beamtime,
only one time-resolved measurement was performed with
the TiCl$_3$-doped NaAlH$_4$ (a8a) sample.
After measuring the resolution at room temperature,
the sample was heated in steps up to 118$^\circ$C,
where the onset of hydrogen release was noticed.
During 20~h, the scattering intensity decayed by more than 50\%.
Before loosing the hydrogen completely,
the temperature was increased to 131, 137, 167, 182$^\circ$C
for rather short measurements.

Fig.~\ref{Fti4-sqw} shows the weak quasielastic scattering at~118$^\circ$C.
The fit consists of a free background, a $\delta$ line,
and a Lorentzian,
with a fixed width of 0.20~$\mu$eV derived from the Arrhenius description
(Fig.~\ref{Fna2-arr}) of Lorentzian~1 of the Na$_3$AlH$_6$ analysis.

\section{Kinetics}\label{Skin}

The results of the spectral analysis shall now be used for
a quantitative description of the time-resolved measurements.
Originally,
one spectrum was saved every 5 minutes.
In our analysis, we binned them into blocks of 20~min.
While 20~min spectra are quite noisy,
it is perfectly possible to perform a full spectral fit.
The well-measured resolution function is taken into account 
just as in the last section.
This fitting which can be completely automatized
must, of course, not be construed as a model validation.
But given a valid model, it is an efficient way of parameter extraction.
The only condition is that fit parameters must not be close to degeneration.
Therefore, it is of paramount importance that 
fixed Lorentzian linewidths
as obtained from the Arrhenius laws of Fig.~\ref{Fna2-arr}
are imposed.

\subsection{\boldmath N\lowercase{a}$_3$A\lowercase{l}H$_6$}

\begin{figure}[thb]
 \centerline{\includegraphics*[width=.9\columnwidth]{\dirfig 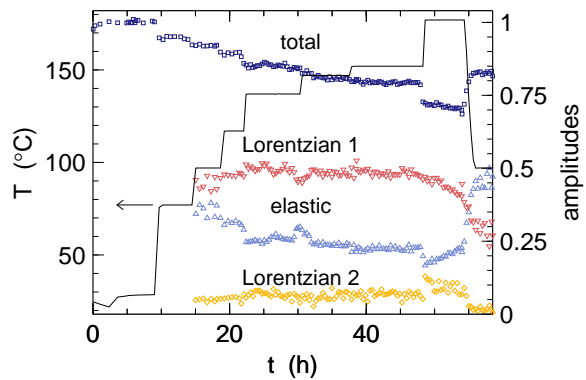}}
 \caption{Time-resolved backscattering measurement of Na$_3$AlH$_6$.
Left scale: Experimental temperature sequence.
Right scale: Total backscattering intensity,
elastic intensity, and Lorentzian amplitudes.}
 \label{Fna2-tsc}
\end{figure}

Our time-resolved measurement of Na$_3$AlH$_6$ is
summarized in Fig.~\ref{Fna2-tsc}.
The applied temperature sequence is shown 
along with the total backscattering intensity~$I$
(the integral of $S(q,\omega)$ over the experimental energy range
$-8.6\ldots8.6$~$\mu$eV) and
the outcome of the fits,
namely, the elastic intensity $a_\delta$
and the two Lorentzian amplitudes $a_{1,2}$.

When heating step by step
from room temperature to 152$^\circ$C,
$I$ and $a_\delta$ exhibit a parallel decrease:
This is, at least qualitatively, the expected
evolution of a Lamb-M\"ossbauer factor.
Reliable amplitudes $a_{1,2}$ can only be determined 
from 97$^\circ$C onwards.
In the range $97\ldots152^\circ$C,
both Lorentzian amplitudes are basically constant,
which validates our fitting method with fixed linewidths~$\Gamma_{1,2}$.

Shortly before heating to 147$^\circ$C,
there is a transient, complementary excursion in $a_\delta$ and
$a_1$, which we cannot explain.
Another strange feature is 
the complementary step in $a_\delta$ and $a_2$ after heating to 177$^\circ$C.
It is related to the outlier $\Gamma_2(177^\circ{\rm C})$
in Fig.~\ref{Fna2-arr}, indicating a limitation of our fit.

We measured about 6~h at 177$^\circ$C.
During this time,
the quasielastic amplitude $a_1$ decreased by 0.14.
This was partly compensated by an increase in elastic scattering of 0.05
so that the total backscattering decreased only by 0.06.

At this point, we decided not to wait for complete decomposition
but to cool back to 117$^\circ$C for comparison with spectra measured
before the excursion to higher temperatures.
Comparison revealed a decrease $\Delta a_1=-0.15$, $\Delta a_2=-0.03$,
partly compensated by an increase $\Delta a_\delta=0.10$,
resulting in a decrease of the total backscattering of $\Delta I=-0.09$.

\subsection{\boldmath N\lowercase{a}A\lowercase{l}H$_4$}

\begin{figure}[thb]
 \centerline{\includegraphics*[width=.9\columnwidth]{\dirfig 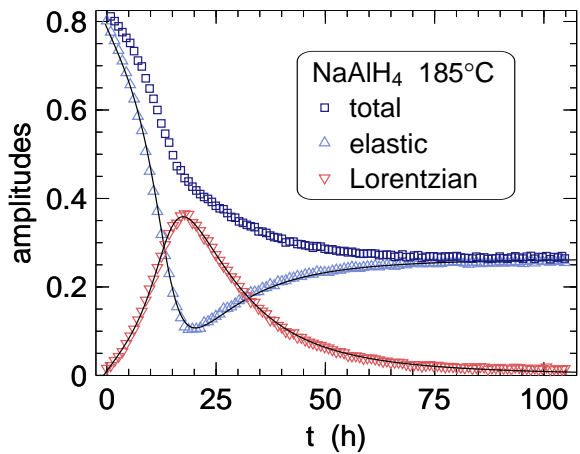}}
 \caption{Transformation of NaAlH$_4$ at 185$^\circ$C:
Total backscattering intensity~$I$
and fitted amplitudes $a_\delta$ and~$a_1$ [Eq.~(\ref{ESqw})] versus time.
Solid lines are fits with a kinetic model described
in Sect.~\ref{Skimo}.}
 \label{Fna4-fitfit}
\end{figure}

As already shown in Fig.~\ref{Fna4-sqw},
quasielastic scattering slowly emerged
after heating NaAlH$_4$ to a temperature 
in the range of 170\ldots185$^\circ$C,
only to disappear some 10~h later.
For a closer analysis,
all 20~min spectra were fitted with a free background,
a $\delta$ line, and one Lorenztian of fixed width $\Gamma_1(T)$
taken from the Arrhenius fit of Fig.~\ref{Fna2-arr}.

Fig.~\ref{Fna4-fitfit} depicts
the total backscattering intensity~$I$
and the amplitudes $a_\delta$,~$a_1$ as function of time,
along with a fit that will be discussed below in Sect.~\ref{Skimo}.
The quasielastic amplitude~$a_1$ reaches a maximum after about 18~h,
before it decays slowly.
In the long term, this decay is exponential,
with a half life of about 10~h.
The elastic intensity~$a_\delta$
decreases to less than 1/6 of its initial value,
before recovering up to about 1/4 of the room temperature value.
Total backscattering decreases continuously.

\begin{figure}[thb]
 \centerline{\includegraphics*[width=.75\columnwidth]{\dirfig 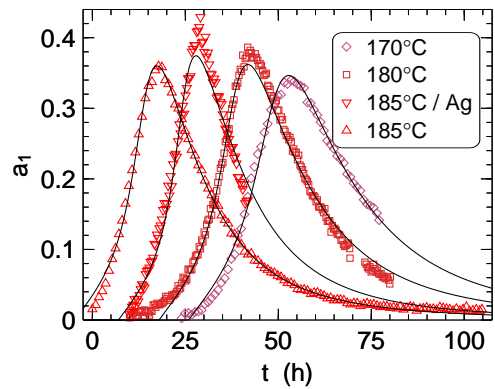}}
 \caption{Quasielastic amplitude~$a_1$
in four experiments on NaAlH$_4$ versus time.
Time scales are arbitrarily shifted to keep the curves separated.
``Ag'' designates a sample that was isolated from its Al container
by a silver foil.
Solid lines are fits with the same kinetic model as in
Fig.~\protect\ref{Fna4-fitfit}.}
 \label{Fna4-4fifi}
\end{figure}

This time-resolved measurement was carried out four times
at three different sample temperatures of 170, 180, and 185$^\circ$C.
Fig.~\ref{Fna4-4fifi} shows the time dependence of~$a_1$
for all four runs.
With increasing temperature, the peaks are somewhat sharper;
in particular, the initial slope is much steeper.
There is some fluctuation in the maximum value of~$a_1$,
but altogether the curves  demonstrate 
a very statisfactory reproducibility of
 measurements and data analysis.

\subsection{\boldmath N\lowercase{a}A\lowercase{l}H$_4$~(a8a)}

\begin{figure}[thb]
 \centerline{\includegraphics*[width=.9\columnwidth]{\dirfig 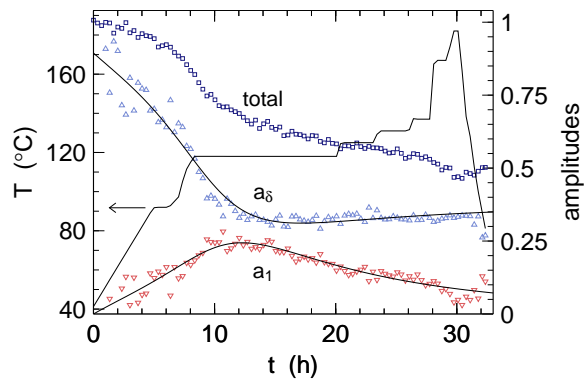}}
 \caption{Time-resolved backscattering measurement
of TiCl$_3$-doped NaAlH$_4$ (a8a).
Left scale: Experimental temperature sequence.
Right scale: Total backscattering intensity,
elastic intensity, and Lorentzian amplitudes.
Solid lines are fits with the same kinetic model as in
Fig.~\protect\ref{Fna4-fitfit},
to be discussed in Sect.~\ref{Skimoti}.}
 \label{Fti4-tsc}
\end{figure}

For the TiCl$_3$-doped NaAlH$_4$ (a8a) sample,
we have so far undertaken one measurement only,
with an improvized, unsystematic temperature sequence,
as shown in Fig.~\ref{Fti4-tsc}.
Quasielastic scattering seems to set in below 80$^\circ$.
However, fits at such low temperatures are not reliable,
because the linewidth $\Gamma_1(T)$ is far below the instrumental resolution.
Once the sample is heated to 118$^\circ$C,
the quasielastic amplitude reaches its maximum within less than 2~h.
Then, a slow decay sets in, with a half life of about~14~h.
The temperature excursion to 167 and 182$^\circ$C is accompanied
by a dip in~$a_1$: 
It is not clear whether this is physical;
it depends critically
on the Arrhenius fit used for imposing~$\Gamma_1(T)$.

\section{Interpretation}\label{Sphys}

\subsection{Quasielastic Scattering by
            \boldmath N\lowercase{a}$_3$A\lowercase{l}H$_6$}

The Lorentzians used to fit the quasielastic scattering
have 
temperature-independent amplitudes $a(q)$
and wavenumber-independent linewidths $\Gamma(T)$
in a good first approximation.
This is the signature of localized processes.
We see no scattering by long-ranged diffusion:
Most probably, diffusion of H$_2$ is too fast
and diffusion of heavier species is too slow to be observed within
the dynamic window of SPHERES.

Lorentzian~1 is due to internal motion of Na$_3$AlH$_6$:
It is dominant in the quasielastic scattering of
a freshly prepared sample,
and found as a transient
during the decomposition of NaAlH$_4$,
as expected from the two-step reaction formula (\ref{Estep1}),~(\ref{Estep2}).

In contrast, Lorentzian~2 only 
is a small contribution to scattering of the fresh Na$_3$AlH$_6$ sample;
it is not visible during decomposition of NaAlH$_4$.
Its origin must be left unresolved.

The observed $q$ dependence of the quasielastic amplitudes $a_{1,2}$,
approximately described by Eq.~(\ref{Eq2sat}),
is not compatible with elementary jump models
that would require an oscillatory $q$ dependence $a\propto[1-j_0(dq)]$
(jump length ${\cal O}(d)$, spherical Bessel function $j_0$) \cite{Bee85}.
One possible explanation could be that each Lorentzian results
from more than one internal mode, 
with similar frequencies, but different jump lengths
so that (\ref{Eq2sat}) results as an average of $1-j_0(dq)$ over several~$d$.

\begin{figure}[thb]
 \centerline{\includegraphics*[width=.75\columnwidth]{\dirfig 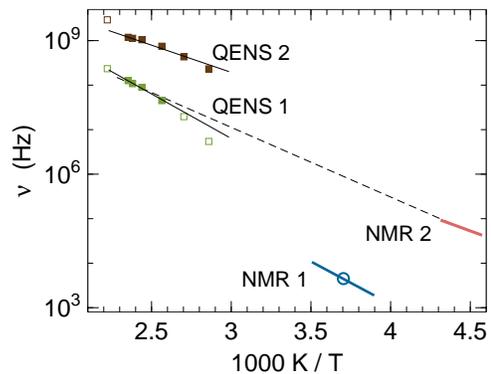}}
 \caption{Rotation frequencies in Na$_3$AlH$_6$ 
as determined by quasielastic neutron scattering (QENS)
and by nuclear magnetic responance (NMR).
The QENS data are the same as in Fig.~\protect\ref{Fna2-arr}.
The data point NMR~1 and the corresponding Arrhenius law
was taken from~\cite{SeVB81}.
The Arrhenius law NMR~2 is from~\cite{VeBE09}.
The dashed line is the same as in Fig.~\ref{Fna2-arr};
it suggests that NMR~2 and QENS~1 are one and the same process.}
 \label{Fna2+nmr}
\end{figure}

Rotational motion in Na$_3$AlH$_6$ below room temperature
was studied previously by NMR \cite{SeVB81,VeBE09}.
Results indicated thermally activated rotational jumps
of AlH$_6$ octahedra around C$_4$ axes.
Jump frequencies, described by Arrhenius laws,
are indicated in Fig.~\ref{Fna2+nmr}.
The slow C$_{4XY}$ rotations of Ref.~\cite{VeBE09}
were found to admit activation energies between 20 and 30~kJ/mol.
Choosing the latter value,
the Arrhenius law (the dashed line in Figs.\ \ref{Fna2-arr} and~\ref{Fna2+nmr})
extrapolates quite well towards the linewidths of our Lorentzian~1.
This further supports our interpretation of the Lorentzian~1 being due
to internal rotations in~Na$_3$AlH$_6$.

\subsection{Decomposition of \boldmath N\lowercase{a}A\lowercase{l}H$_4$}
 \label{Skimo}

The increase and decrease of quasielastic scattering
(Figs.~\ref{Fna4-sqw}, \ref{Fna4-fitfit})
reflect the production and consumption of Na$_3$AlH$_6$
in the two-step reaction (\ref{Estep1}),~(\ref{Estep2}), respectively.
For better readability, we abbreviate these reaction equations as
$A\to B\to C$,
where $A$ stands for NaAlH$_4$,
$B$ for 1/3 Na$_3$AlH$_6$,
and $C$ for NaH.
Side products of the reactions are not denoted,
because they are irrelevant for the observed spectra:
Al has a negligible neutron cross section,
and H$_2$ leaves the sample so rapidly
that it does not contribute to the scattering either.

Concentrations shall be written as dimensionless mol/mol fractions,
normalized to a pure NaAlH$_4$ sample.
Accordingly, the initial concentrations are $[A]=1$, $[B]=[C]=0$.
Since our scattering signal is almost exclusively due to bound hydrogen,
we set
\begin{eqnarray}
 \label{Eka1}  a_1(t) &=& \frac{1}{2} [B]f_B,\\
 \label{Ekad}  a_\delta(t) &=& [A]f_A+\frac{1}{4}[C]f_C,
\end{eqnarray}
where the $f$ are Lamb-M\"ossbauer factors.

On this base, we searched for rate equations that reproduce
the observed time series $a_1(t)$ and~$a_\delta(t)$.
We first used the code generator \textit{kinpy}~\cite{kinpy}
to try kinetic models based on stochiometric reaction equations.
It turned out that the relatively sharp peak in $a_1(t)$
can only be reproduced if some autocatalysis is assumed.
To improve the agreement with the experimental $a_\delta(t)$ and~$a_1(t)$,
we admitted concentration dependences
with fractional exponents.
In this heuristic way,
we found that the following kinetic model is about the simplest one
that is compatible with the measured $a_\delta(t)$ and~$a_1(t)$:
\begin{equation}\label{Ekimo}
\begin{array}{lcl}
{\rm d}[A]/{\rm d}t &=& - k_{00}[A] - k_{01}[A][B]^{2},\\[1.8ex]
{\rm d}[B]/{\rm d}t &=& -{\rm d}[A]/{\rm d}t-{\rm d}[C]/{\rm d}t,\\[1.8ex]
{\rm d}[C]/{\rm d}t &=& k_{10}[B]^{4/3} + k_{11}[B]^{4/3}[C]^{2/3},\\
\end{array}
\end{equation}
At 185$^\circ$C,
the following rate coefficients are found:
\begin{equation}\label{Ekico}
\begin{array}{lcl}
  k_{00} &=& 0.029~\mbox{h}^{-1},\\
  k_{01} &=& 0.55~\mbox{h}^{-1},\\
  k_{10} &=& 0.016~\mbox{h}^{-1},\\
  k_{11} &=& 0.097~\mbox{h}^{-1}.
\end{array}
\end{equation}
The spontaneous rate coefficients $k_{j0}$ ($j=0,1$)
are smaller by about an order of magnitude
than the autocatalytic ones $k_{j1}$,
and the second autocatalytic step is much slower than the first one,
as was to be expected from the shape of $a_1(t)$.

For the fits at different temperatures in Fig.~\ref{Fna4-4fifi},
the spontaneous rates $k_{j0}$ were fixed.
Besides a trivial time shift, the only adjustable parameters were
the two autocatalytic rates $k_{j1}$.
Their fitted values show a plausible, weak temperature dependence.
The fits are not perfect,
but this does does not necessarily indicate a failure of our kinetic model;
it could also point to the limits of experimental reproducibility
(sample preparation, thermal history).

The exponents in~(\ref{Ekimo}) cannot be determined very accurately;
it can only be said that they are compatible with multiples of 2/3,
as expected for an autocatalytic reaction that takes place at a surface.
We suggest that $k_{j0}$ stands for nucleation
and $k_{j1}$ for crystallite growth.

Our model is perfectly compatible with the reaction pathways
proposed in a density-functional study \cite{GuHO08}.
Based on first-principles calculations,
the migrating species were suggested to be
AlH$_3$ and NaH vacancies.
The lowest activation energy was found for AlH$_3$ vacancies,
leading to the following pathway for reaction~(\ref{Estep1}):
\begin{equation}
 n\mbox{NaAlH$_4$} \rightarrow
     n\mbox{NaAlH$_4^{\text{AlH$_3$}}$} + \mbox{Al}
      + \textstyle\frac{3}{2} \mbox{H}_2,   
\end{equation}
where the superscript denotes one vacancy,
and $n$ indicates an arbitrary amount of bulk material.
The proposed diffusion mechanism
 also involves an AlH$_5^{2-}$ ion.
Anyway, the vacancy ultimately reaches a 
NaAlH$_4$--Na$_3$AlH$_6$ boundary,
where it annihilates, releasing
an excess Na$^{+}$ that aggregates with the growing Na$_3$AlH$_6$ phase.
From another computational study we learn that this growth
additionally requires a two-step transformation
 AlH$_4^{-}$ $\to$ AlH$_5^{2-}$ $\to$ AlH$_6^{3-}$ \cite{DaMa10}.

Coming back to Ref.~\cite{GuHO08},
an alternate pathway starts with the unassisted release of
Na$^{+}$ and H$^{-}$ at the boundary of the growing Na$_3$AlH$_6$
according to
\begin{equation}
 n\mbox{NaAlH$_4$} \rightarrow
     (n-1)\mbox{NaAlH$_4^{2\text{NaH}}$} + \mbox{Na$_3$AlH$_6$}.
\end{equation}
The two ionic vacancies migrate together to the Al--NaAlH$_4$ boundary,
where hydrogen is released:
\begin{equation}
 n\mbox{NaAlH$_4^{\text{NaH}}$} \rightarrow
    (n-1)\mbox{NaAlH$_4$} + \mbox{Al} + \textstyle\frac{3}{2}\mbox{H}_2.
\end{equation}
Using positron annilation, it was confirmed experimentally
that vacancies are formed in NaAlH$_4$ during dehydrogenation \cite{SaNA10}.

In these pathways, boundaries play a crucial role:
The NaAlH$_4$--Na$_3$AlH$_6$ boundary as a sink for AlH$_3$ vacancies
and as a source for NaH vacancies,
and the Al--NaAlH$_4$ boundary as a sink for NaH vacancies and as the
location of H$_2$ release.
At least in undoped material,
hydrogen release rates are limited by processes at these boundaries.
This is in accordance with the autocatalytic terms of our kinetic model.

\subsection{Effect of Doping} \label{Skimoti}

In Fig.~\ref{Fti4-tsc},
the kinetic model~(\ref{Ekimo})
was applied to TiCl$_3$-doped NaAlH$_4$,
cycled eight times under hydrogen and quenched in the hydrogenated state.
These fits were at best a very rough first approximation,
because constant rate coefficients were assumed
although the sample temperature varied between 
40 and 137$^\circ$C (the excursion to 167\ldots182$^\circ$C
was excluded from the fit).
The model may even be entirely inappropriate,
because the rate-limiting processes
in doped alanate could be qualitatively different from those in pure material.
According to~\cite{GuHO08},
hydrogen release in pure material is limited by surface processes,
whereas in doped material rates are determined by vacancy diffusion.

Under these reservation,
the coefficients are about
\begin{equation}\label{Ekicoti}
\begin{array}{lcl}
  k_{00} &=& 0.05~\mbox{h}^{-1},\\
  k_{01} &=& 0.8~\mbox{h}^{-1},\\
  k_{10} &=& 0.12~\mbox{h}^{-1};\\
\end{array}
\end{equation}
$k_{11}$ had no noticeable influence upon the time dependence
of $a_\delta(t)$ and~$a_1(t)$.
All three values in (\ref{Ekicoti}) are higher than their
counterparts in (\ref{Ekico}) although the latter
were determined at the considerably higher temperature of~170$^\circ$C.
Doping was expected to increase the coefficient $k_{00}$,
so that the first decomposition step starts at much lower temperatures
than in the undoped material.
However, the effect of doping on the second decomposition step
seems to be even stronger: The coefficient $k_{10}$ is larger
than in the undoped 185$^\circ$C data by a full order of magnitude.
This correlates with the observation 
that the peak height of $a_1(t)$
is much lower in the doped
than in the undoped material.

\section{Conclusion}\label{Scon}

The high flux, low background, and fine resolution of the
new backscattering spectrometer SPHERES opens up new perspectives
for time-resolved studies of weak and narrow quasielastic spectra.
The present work reveals the potential of this method
for investigations of reaction kinetics.

The desorption of hydrogen from sodium alanate is an almost
perfect test case,
because on the one hand with its five different reactants
it is complex enough to be interesting,
while on the other hand the neutron spectra are not too complex,
because two of the reactants do not noticeably contribute to the scattering,
and only one of them, the intermediate reaction product Na$_3$AlH$_6$,
shows quasielastic scattering.
Once these quasielastic spectra are well characterized,
the time-resolved measurements allow for an automatic data reduction
that yields a small number of time-dependent amplitudes.
These amplitudes can be related in a straightforward way
[Eqs. (\ref{Eka1}), (\ref{Ekad})] to reactant concentrations.

A time-resolved measurement of NaAlH$_4$ decomposition
at 170\ldots185$^\circ$C
was fitted by a physically meaningful set of
reaction rate equations.
According to this model,
both desorption steps (\ref{Estep1}),~(\ref{Estep2})
are controlled by nucleation and growth.

Next, it will be interesting to investigate more systematically
the temperature dependence of reaction rates.
To initiate the decomposition of NaAlH$_4$,
there seems to be only a small temperature range between about 165$^\circ$C
and the melting point.
However, once the reaction has started,
its autocatalytic sequel possibly remains active at considerably lower
temperatures.

While special care was taken to produce all samples
in the same manner,
it would be interesting to investigate how the kinetics depends
on the initial texture
and on the hydrogen loading/unloading prehistory.

Most importantly,
the basis is now available for systematically investigating 
how the single reaction steps are modified by the presence of a dopant.
As a very preliminary result,
we hypothize that TiCl$_3$ mainly accelerates the spontaneous decomposition
of the intermediate reactant Na$_3$AlH$_6$.
However, this still needs thorough verification.
Unfortunately,
only a very narrow quasielastic signal can be observed
at the low reaction temperatures made possible by doping
(Fig.~\ref{Fti4-sqw}).
Nevertheless,
Fig.~\ref{Fti4-tsc} suggests that a quantitative study of reaction kinetics
will be possible if more systematic time runs will be measured.

\section{Acknowledgements}

We thank Dr.\ Maximilian Fichtner for giving us the possibility 
 to use the energy storage laboratory equipment
for the preparation of the samples. 


\end{document}